%% file: main.tex
\pdfoutput=1
\PassOptionsToPackage{table}{xcolor}
\documentclass[sigconf]{acmart}
\usepackage{indentfirst}

\copyrightyear{2026}
\acmYear{2026}
\setcopyright{cc}
\setcctype{by}
\acmConference[ICCAD '26]{IEEE/ACM International Conference on Computer-Aided Design}{November 08--12, 2026}{San Jose, CA, USA}
\acmBooktitle{IEEE/ACM International Conference on Computer-Aided Design (ICCAD '26), November 08--12, 2026, San Jose, CA, USA}
\acmDOI{10.1145/3831252.3834108}
\acmISBN{979-8-4007-2873-0/2026/11}

\usepackage{algpseudocode}
\algrenewcommand\algorithmicrequire{\textbf{Input:}}
\algrenewcommand\algorithmicensure{\textbf{Output:}}
\usepackage{algorithm}
\usepackage{graphicx}
\usepackage{textcomp}
\usepackage{makecell}
\usepackage{caption}
\usepackage{tabularx}
\usepackage{booktabs}
\usepackage{siunitx}
\usepackage{comment}
\usepackage{multirow}
\usepackage{colortbl}
\usepackage{threeparttable}
\newcommand{\runtimemark}{\textsuperscript{\(\lozenge\)}}
\usepackage{array}
\setcitestyle{numbers,sort&compress}
\usepackage{placeins}

\begin{document}
\renewcommand\tabularxcolumn[1]{m{#1}}

\title[\texttt{Forbench}: Symbolic Simulation Helps  Make Your Test\underline{bench} More \underline{For}mal]{\texttt{Forbench}: Symbolic Simulation Helps  Make Your \\ Test\underline{bench} More \underline{For}mal}


\author{%
  {\normalsize
  \begin{tabular}{c}
    Ziyi Yang*\textsuperscript{1}, 
    Wenbin Che*\textsuperscript{1}, 
    Ziyue Zheng\textsuperscript{1}, 
    Guangyu Hu\textsuperscript{2}, 
    Hongce Zhang\textsuperscript{1,2,\textdagger}\\[2pt]
    {\normalfont\textsuperscript{1}The Hong Kong University of Science and Technology (Guangzhou)}\\[-1pt]
    {\normalfont\textsuperscript{2}The Hong Kong University of Science and Technology} \\
    {\normalfont\normalsize{\{zyang957, wche691, zzheng989\}@connect.hkust-gz.edu.cn}, ghuae@connect.ust.hk, hongcezh@hkust-gz.edu.cn}
  \end{tabular}}
  }
\makeatletter
\g@addto@macro\thankses{\thanks{*These authors contributed equally.}}
\g@addto@macro\thankses{\thanks{\textdagger Corresponding author.}}
\makeatother
\gdef\authors{Ziyi Yang, Wenbin Che, Ziyue Zheng, Guangyu Hu, and Hongce Zhang}
  






\renewcommand{\shortauthors}{Yang et al.}

\begin{abstract}
\input{sections/abstract}

\end{abstract}





\maketitle

\newcommand{\wenbin}[1]{\textcolor{blue}{[\textbf{wenbin}: #1]}}
\newcommand{\hongce}[1]{\textcolor{red}{[\textbf{hongce}: #1]}}
\newcommand{\ziyi}[1]{\textcolor{green}{[\textbf{ziyi}: #1]}}

\definecolor{lightblue}{RGB}{223, 235, 247}

\section{Introduction}
\input{sections/introduction}

\section{Challenges of Property Specification: A Motivating Example}
\input{sections/motivation_new}

\section{Related Work: Existing Symbolic Methods}
\input{sections/related}

\section{The \texttt{Forbench} Methodology}
\input{sections/method}

\section{Experiment}
\input{sections/experiment}

\vspace{-0.5em}
\section{Conclusion}

\input{sections/conclusion}

\begin{acks}
This work is supported by the National Natural Science Foundation of China (grant no. 62304194)
\end{acks}

\bibliographystyle{unsrtnat}
\bibliography{ref}

\end{document}

%% file: sections/abstract.tex
Simulation remains the dominant approach in pre-silicon verification due to its ease of deployment and intuitive workflow. 
However, as simulation only explores a limited subset of possible execution traces within feasible time budgets, it often fails to explore rare corner cases, leaving latent bugs undetected. 
%
In contrast, formal verification 
offers mathematically rigorous guarantees of correctness. However, its practical adoption is constrained, not only by the scalability challenges over large-scale designs, but also by the change of mindset from stimulus-driven operations to the sequence-centric axiomatic view of design behaviors, introducing extra difficulty of writing precise properties to capture the exact verification intent.


This paper aims to lower the barrier of applying formal methods in verification, by making simulation ``more formal.'' 
It introduces \texttt{Forbench}, a word-level symbolic simulation framework that retains the familiar execution semantics of simulation but augments it with solver-backed symbolic signals and state transitions, enabling systematic exploration of RTL behaviors under symbolic inputs and conditions. 
It offers a Python interface, similar to the existing simulation-based frameworks, for defining constraints, coordinating symbolic (co-)simulations, and performing property checks.
In additional to this more accessible interface, 
experiments also show that \texttt{Forbench} achieves notably speed-up over prior symbolic methods without the loss of coverage. 


%% file: sections/introduction.tex

Functional verification is fundamental to 
ensuring design correctness. 
As design scales continue to grow, the engineering effort of verification has also been rising accordingly,
demanding longer development cycles and substantial human resources---often rivaling or even exceeding those devoted to the design process itself~\cite{FosterWRG2020Part4}.
 Consequently, verification has become a primary bottleneck in the development flow~\cite{FosterWRG2022Part8}, underscoring the urgent need for methodologies that reduce manual effort, enhance coverage, and deliver stronger guarantees of correctness under constrained time and resource budgets.

\input{table/overview.tex}

Simulation-based verification remains the dominant industrial practice, valued for its intuitive operational execution model and low expertise requirement. Engineers construct executable testbenches that generate concrete stimuli, observe responses, and compare them against expected behaviors. Recent advances, such as coverage-guided and constraint-random testing~\cite{laeufer2018rfuzz,hur2021difuzzrtl,trippel2022fuzzing}, have improved coverage by automating and directing stimulus generation, yet these methods fundamentally explore only a fraction of the input space. Each test run exercises a single execution trace, offering limited confidence rather than guarantees of correctness. While simulation offers rapid feedback and scalability, its limited coverage leaves subtle corner cases unchecked, which may later manifest as a costly bug escape.\looseness=-1

By contrast, formal property verification (FPV) seeks \textit{exhaustive assurance}.
It describes verification intent using property specification languages, such as SystemVerilog Assertions (SVAs), and employs model checking~\cite{biere1999symbolic, een2011efficient, clarke1986automatic, clarke1997model, clarke2018handbook} to mathematically prove or refute the properties.
Unlike simulation, FPV does not require engineers to manually construct stimulus sequences, as the solver implicitly explores all admissible input behaviors, thereby reducing the risk of untested scenarios.

Despite these advantages, adopting FPV in practice demands substantial expertise and a shift in verification mindset. Engineers must move from procedural, stimulus-driven thinking to an axiomatic, property-centric view of system behavior. This transition entails not only the knowledge of formal methods, but also the fluency in assertion semantics.
Being axiomatic, SVA is generally more difficult to express procedural behaviors, such as the transactions needed to interact with the design under verification, and it is difficult to compose complex properties from simpler ones. Therefore, crafting precise SVAs that faithfully capture verification intent remains a nontrivial task (an example is given in Section~\ref{sec:motivation_new}). Moreover, when properties remain inconclusive after hours of runtime, it lacks a straightforward mechanism to control the trade-off between computational complexity and design coverage.
Altogether, these challenges call for a better interface to integrate formal techniques in functional verification.

As an effort to make formal verification more accessible, 
this paper introduces \texttt{Forbench}, a formal testbench paradigm that offers a third path between simulation and FPV as shown by the comparison in  Table~\ref{tab:overview}.
\texttt{Forbench} has the following key features:
\begin{itemize}
    \item It blurs the boundary between the two existing methodologies as it preserves the operational mindset, using testbench structure familiar to simulation engineers, while incorporating symbolic reasoning to achieve deeper and more systematic coverage. This unique feature enables engineers to express verification intent in a more natural and procedural style while still enjoying the benefit of formal reasoning.
    \item Furthermore, the degree of symbolic exploration is easily tunable via choices of being concrete or symbolic for each individual value in the stimuli, allowing adjusting verification hardness with minimal effort.
    \item To further improve usability, \texttt{Forbench} provides a Python interface and supports two complementary testbench styles (active-stepping and coroutine-based), making it readily adaptable to different verification tasks and user preferences.
\end{itemize}

By integrating these capabilities, \texttt{Forbench} leverages the strengths of both simulation and formal methods, 
paving the way toward more scalable and accessible formal reasoning in industrial hardware verification.\looseness=--1

The rest of the paper is organized as follows. Section~\ref{sec:motivation_new} presents an example that 
illustrates the hurdle of property specification in the current FPV methodology. Section~\ref{sec:related} discusses the limitation of other formal/semi-formal symbolic methods, motivating the need for \texttt{Forbench}. Section~\ref{sec:method} introduces the idea of formal testbenches and the core technologies that enable \texttt{Forbench}. 
Section~\ref{sec:experiment} evaluates the performance of \texttt{Forbench} symbolic simulation against other verification methodologies. Finally, Section~\ref{sec:conclusion} concludes the paper.

%% file: table/overview.tex
\newcolumntype{Y}{>{\centering\arraybackslash}X}
\begin{table}[t]
  \centering
  \caption{Comparison of different verification methodologies. \texttt{Forbench} strikes a balance between formal methods and simulation-based approaches.}
  \vspace{-1em}
  \label{tab:overview}
  \begin{threeparttable}
    \setlength{\tabcolsep}{5pt}
    \renewcommand{\arraystretch}{1.1}
    \begin{tabularx}{\columnwidth}{
  @{}
  >{\raggedright\bfseries\arraybackslash}m{0.25\columnwidth}
  @{\hspace{2pt}}
  >{\centering\arraybackslash}m{0.25\columnwidth}  
  >{\centering\arraybackslash}m{0.20\columnwidth}  
  >{\centering\arraybackslash}m{0.18\columnwidth}  
  @{}
}
      \toprule
      & \makecell[c]{\textbf{Simulation}\\ \cite{guo2025gem,hur2021difuzzrtl,laeufer2018rfuzz,trippel2022fuzzing}}
      & \makecell[c]{\textbf{FPV\tnote{\ddag}}\\ \cite{xiao2023fuzzbtor2,fang2023wasim,su2025deeply}}
      & \makecell[c]{\cellcolor{lightblue}\textbf{Forbench}\\ \cellcolor{lightblue}\textbf{(Ours)}} \\

      \midrule
      Expertise        & Low                 & High              & \textbf{Medium} \\
      Artifact         & Testbench           & Assertions        & \textbf{Testbench} \\
      Stimulus setup   & Concrete            & Not needed       & \textbf{Symbolic} \\
      Runtime          & Low\runtimemark     & High              & \textbf{Tunable} \\
      \small Execution model  & Operational         & Axiomatic         & \textbf{Operational} \\
      Coverage         & Limited             & High              & \textbf{High} \\
      \bottomrule
    \end{tabularx}
    \begin{tablenotes}[flushleft]\footnotesize
      \item[\ddag] FPV: Formal Property Verification.
      \item[\(\lozenge\)] Low runtime for a single concrete run; high time to achieve high coverage.
    \end{tablenotes}
  \end{threeparttable}
\vspace{-2.5em}
\end{table}

%% file: sections/motivation_new.tex
\label{sec:motivation_new}

\begin{figure}[t!]
  \centering
  \includegraphics[width=\columnwidth]{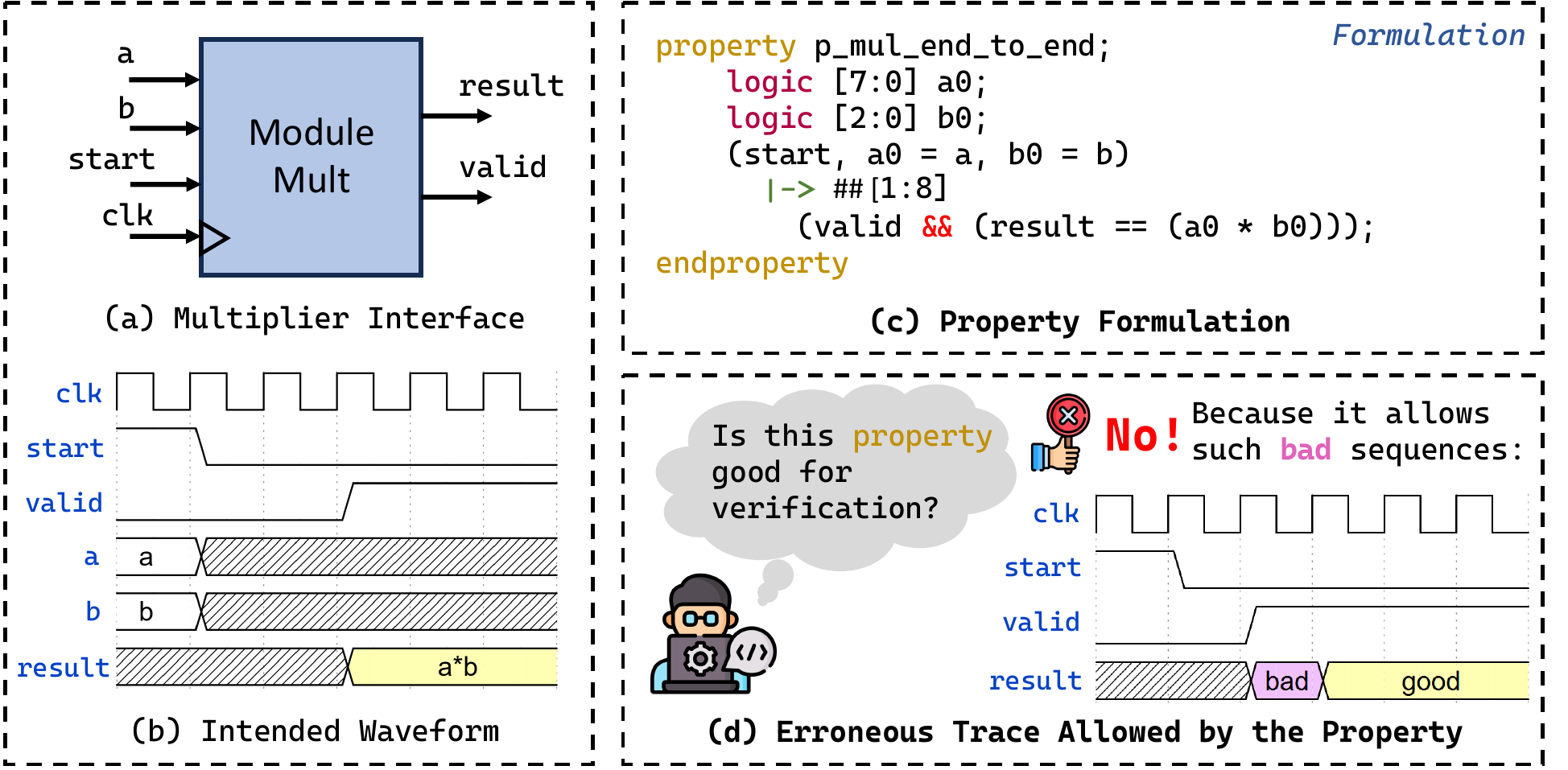}
  \vspace{-1em}
  \caption{A motivating example showing how the intuitive SVA formulations may fail to precisely capture the verification intent. (a) and (b) illustrate the intended interface and the expected correct behavior. (c) shows a natural property formulation for checking this intent. However, as illustrated in (d), this formulation may still accept an erroneous trace.} 
  \label{fig:motivation_new}
  \vspace{-1em}
\end{figure}

In this section, we use the formal property verification of a multiplier as a motivating example. The interface of the multiplier is illustrated in Fig.~\ref{fig:motivation_new}(a). The design accepts inputs \texttt{a} (multiplicand) and \texttt{b} (multiplier), and produces the output \texttt{result}. A control signal \texttt{start} indicates the initiation of computation: when asserted, the module samples the values of \texttt{a} and \texttt{b} in the same cycle and begins execution. Upon completion, the signal \texttt{valid} is asserted, indicating that \texttt{result} holds the final multiplication outcome. The computation latency ranges from 1 to 8 cycles, and the intended input/output interaction is depicted in Fig.~\ref{fig:motivation_new}(b).

To verify the multiplier follows the intended behavior using formal property verification, one may naturally construct a property such as the one shown in Fig.~\ref{fig:motivation_new}(c). This property introduces local variables \texttt{a0} and \texttt{b0} to capture the input operands when \texttt{start} is asserted. The expectation is that, once computation is triggered, the signal \texttt{valid} will be asserted together with the correct result within 1 to 8 cycles. This intent is encoded using the implication operator \texttt{|->} and the bounded delay \texttt{\#\#[1:8]}.

However, this seemingly reasonable property is flawed in practice. In particular, it admits erroneous behaviors such as the trace shown in Fig.~\ref{fig:motivation_new}(d), where \texttt{valid} is asserted one cycle earlier than the correct result is produced. Despite this mismatch, the property still passes: the temporal expression \texttt{|-> \#\#[1:8]} only requires that there exists at least one cycle within the specified window where both \texttt{valid} and the correct \texttt{result} coincide. As a consequence, premature presence of \texttt{valid} is not ruled out, as long as a correct pairing later occurs. This is problematic in realistic when downstream modules consume \texttt{result} immediately upon observing \texttt{valid}, thereby propagating incorrect data. Notably, replacing the conjunction \texttt{\&\&} (marked in red in the figure) with an implication \texttt{|->} does not resolve this issue, as the underlying temporal permissiveness remains unchanged.

The situation could become even more complicated when one needs to incorporate additional protocol constraints. For example, if we wish to also enforce that \texttt{start} must not be re-asserted while a computation is already in progress when the I/O protocol of this module does not allow overlapping execution intervals. This typically requires intricate sequence constructions involving overlap control and likely also repetition operators. Thus, the user of SVA must carefully navigate subtle semantic distinctions, such as \texttt{within} vs. \texttt{throughout}, \texttt{until} vs. \texttt{until\_with}, as well as the \emph{three} different types of repetition in SVA. These subtleties significantly increase the difficulty level for a verification engineer to write correct and precise property specifications.\looseness=-1

\begin{figure*}[t!]
  \centering
  \includegraphics[width=0.95\linewidth]{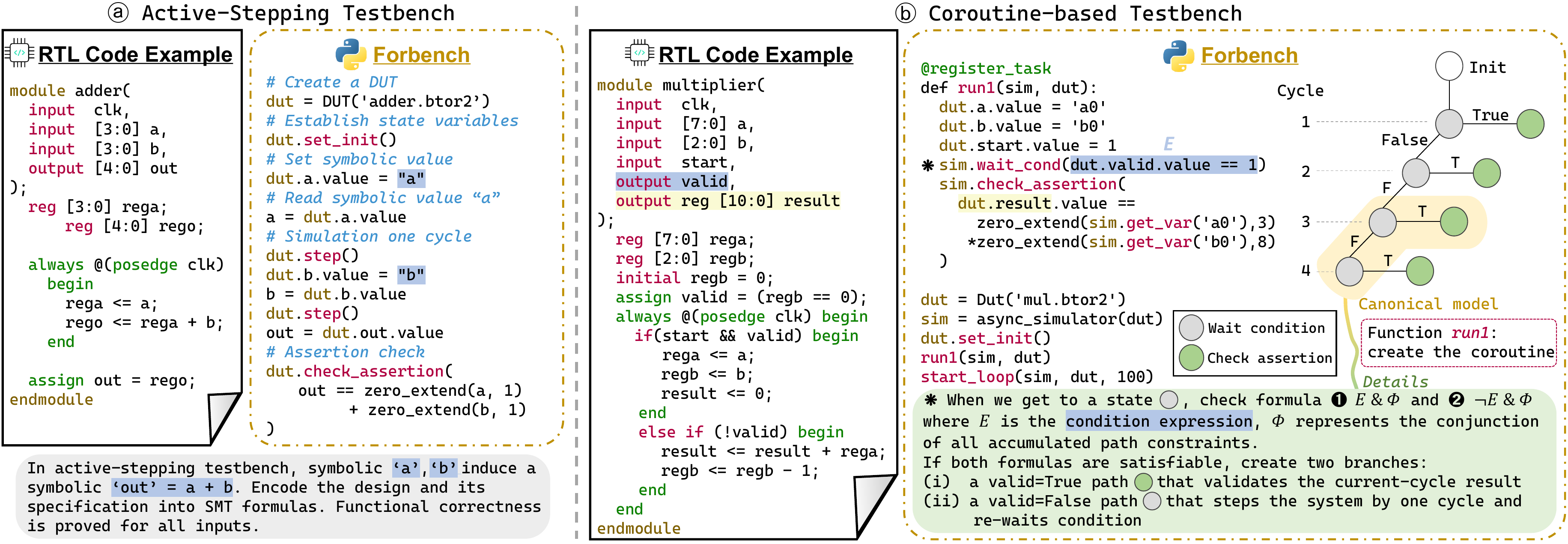}
  \caption{A demonstration of the \texttt{Forbench} verification methodology and the two testbench styles. Example~\textcircled{a} on the left uses the active-stepping style, which explicitly invokes ``\textit{dut.step()}'' to advance to the next clock cycle. Example~\textcircled{b} on the right is coroutine-based, which allows concurrently interfacing with multiple DUT interfaces via separate coroutines. }
  \label{fig:interface}
  \vspace{-0.5cm}
\end{figure*}

More broadly, as verification scenarios grow in complexity, SVA provides limited support for hierarchical construction. In contrast to simulation-based verification, where complex operations can be composed from simpler tasks and functions in a modular and reusable manner, in SVA sequence composition is less intuitive. 
Minor mistakes when handling sequences with potential overlaps can easily introduce vacuous satisfaction or over-/under-constrained conditions,  and in the worse case lead to false passes and bug escapes. \looseness=-1 

As a result, even for a simple design such as the multiplier in this example, writing correct and robust properties can be challenging. This motivates the need of a different verification interface to lower the barrier of applying formal methods in functional verification.

%% file: sections/related.tex
\label{sec:related}


In response to the property specification and scalability challenges of FPV, several other symbolic techniques have been proposed to support formal or semi-formal verification of RTL designs.

\subsection{Symbolic Execution}

Symbolic execution~\cite{baldoni2018survey, avgerinos2014enhancing} 
is a program analysis technique that treats inputs as symbolic values and explores program behaviors across multiple execution paths. While it shares the idea of reasoning over symbolic variables, symbolic execution differs fundamentally from symbolic \textit{simulation}:
the former forks execution when encountering conditional branches in the design, whereas the latter maintains a unified symbolic expression that encodes all possible outcomes of those conditions.
For example, when an \texttt{if} statement appears in the program, symbolic execution would create two separate paths to explore the corresponding execution flows, while symbolic simulation represents both possibilities within a single symbolic expression, encoded using, for example, the \texttt{ITE}-operator in Satisfiability Modulo Theories (SMT)~\cite{barrett2018satisfiability}. Therefore, symbolic simulation is less prone to the so-called ``path explosion'' problem~\cite{baldoni2018survey}.

Currently, there is no ready-to-use symbolic execution framework that operates directly on RTL. Prior work such as~\cite{bruns2023processor} instead translates Verilog RTL into C/C++ models using Verilator~\cite{snyder2004verilator}, and then applies a software symbolic execution engine such as KLEE~\cite{cadar2008klee}. This approach has several drawbacks. First, it must bridge the semantic gap between hardware and software execution models, which weakens the correspondence between the original RTL description and the generated software model. Therefore, additional processing is often required to map RTL testbenches to software code and map software-level counterexamples back to the RTL model. Second, the workflow typically involves multiple tools across both hardware and software and lacks a streamlined verification environment familiar to hardware developers. Third, the path-exploration strategy of symbolic execution can suffer from severe path explosion when applied to complex RTL designs.

\subsection{Symbolic Trajectory Evaluation}

Symbolic trajectory evaluation (STE)~\cite{hazelhurst2005symbolic} can be regarded as a form of symbolic simulation, which has been used in the verification of modern processors~\cite{replaceing2009}. 
Classical STE performs formal reasoning mainly using Binary Decision Diagrams (BDDs), which could scale poorly for datapath-heavy designs due to variable-ordering sensitivity and exponential blow-ups. 
Moreover, STE requires the user to be fluent in a domain-specific functional programming language (\texttt{fl}) that could impose a steep learning curve (we argue that today even a notable number of software developers are unfamiliar with functional programming languages, let alone hardware developers), therefore lacking the support for constructing symbolic testbenches in a natural and procedural manner. Apart from the user interface, advanced capabilities such as GSTE~\cite{1218209}  are not openly available in VossII (the open-source STE implementation)~\cite{Seger2020VossII}, therefore, further limiting its practical adoption.


These limitations motivate an alternative symbolic simulation framework. In this paper, we present \texttt{Forbench}, which adopts an SMT-based symbolic simulation approach to improve scalability, particularly for datapath-heavy designs. Furthermore, \texttt{Forbench} provides a high-level Python interface that enables users to construct symbolic testbenches in a familiar procedural style, aligning more naturally with existing simulation-based verification practices.
We expect \texttt{Forbench} to improve both human productivity in setting up verification tasks and also the solving efficiency during the verification process.\looseness=-1

%% file: sections/method.tex
\label{sec:method}

In this paper, we use \texttt{Forbench} to interchangeably refer to (1) the verification methodology involving a Python symbolic testbench and (2) the underlying framework that supports this methodology when the context is clear.
%
%
Section~\ref{3.1} will introduce the overall symbolic testbench paradigm in \texttt{Forbench}.
Sections~\ref{3.2},~\ref{3.3}, and~\ref{3.4} then describe the theoretical foundations and internal implementation of \texttt{Forbench}, followed by discussions of its applications and usage scenarios in Section~\ref{3.5}. 




\subsection{The Symbolic Testbench}
\label{3.1}

\texttt{Forbench} aims to mimic the setup of simulation-based verification while offering symbolic reasoning capabilities.
In concrete RTL simulation, 
there already exist such two styles: active-stepping testbench and coroutine-based testbench.
In the former case, the verification engineer writes a testbench with a single execution flow and the DUT must be explicitly driven by the clock-tick operation in the testbench. This is the case, for example, in the simulation framework \texttt{Picker}~\cite{picker2025}, which also supports free usage of different programming languages to write the testbench. 
%
On the other hand, the coroutine-based simulation framework (such as \texttt{cocotb}~\cite{rosser2018cocotb}) takes a different approach. It mimics the concurrency model of Verilog, which allows concurrent execution of multiple blocks. This is usually more favorable when the user needs to work with multiple DUT interfaces, where the handling of these different interfaces could be coded separately, as if they are put in individual ``always'' blocks in Verilog. In this case, it relies on the simulator to schedule and interleave the operations in different blocks.

As a symbolic simulation framework, \texttt{Forbench} supports \emph{both} styles, as demonstrated by the examples in Fig.~\ref{fig:interface}.  On the left, Example~\textcircled{a} shows the verification of a fixed-latency registered adder using the active-stepping style. Note that the testbench explicitly invokes ``\textit{dut.step()}'' to advance to the next clock cycle. Here, the usage of symbolic value ``a'' and ``b'' enables full-coverage verification of the DUT functionality without exhaustive enumeration of concrete values. Example~\textcircled{b} on the right showcases a coroutine-based testbench verifying a multiplier implemented using iterative additions. Note that the latency of this DUT depends on the input value range and the \texttt{Forbench} testbench uses ``\textit{wait\_cond}'' to achieve---\emph{in the context of symbolic simulation}---the effect of ``\textit{@(posedge dut.valid)}'' in a concrete Verilog simulation, which  waits for the ``valid'' signal to rise. As symbolic simulation allows all possible input ranges, at a certain cycle, this signal may or may not evaluate to logic true. Therefore, ``\textit{wait\_cond}'' involves forking the current coroutine to account for the situations of different latencies.  
More detailed explanations of the Python APIs underlying these two testbench styles are provided in what follows.


\subsubsection{\textbf{Active-stepping testbench}}

\texttt{Forbench} provides a set of Python API that follows the traditional active-stepping simulation workflows while adding solver-backed symbolic simulation and property checking functionalities, including:

\begin{figure}[t!]
  \centering
  \includegraphics[width=0.9\columnwidth]{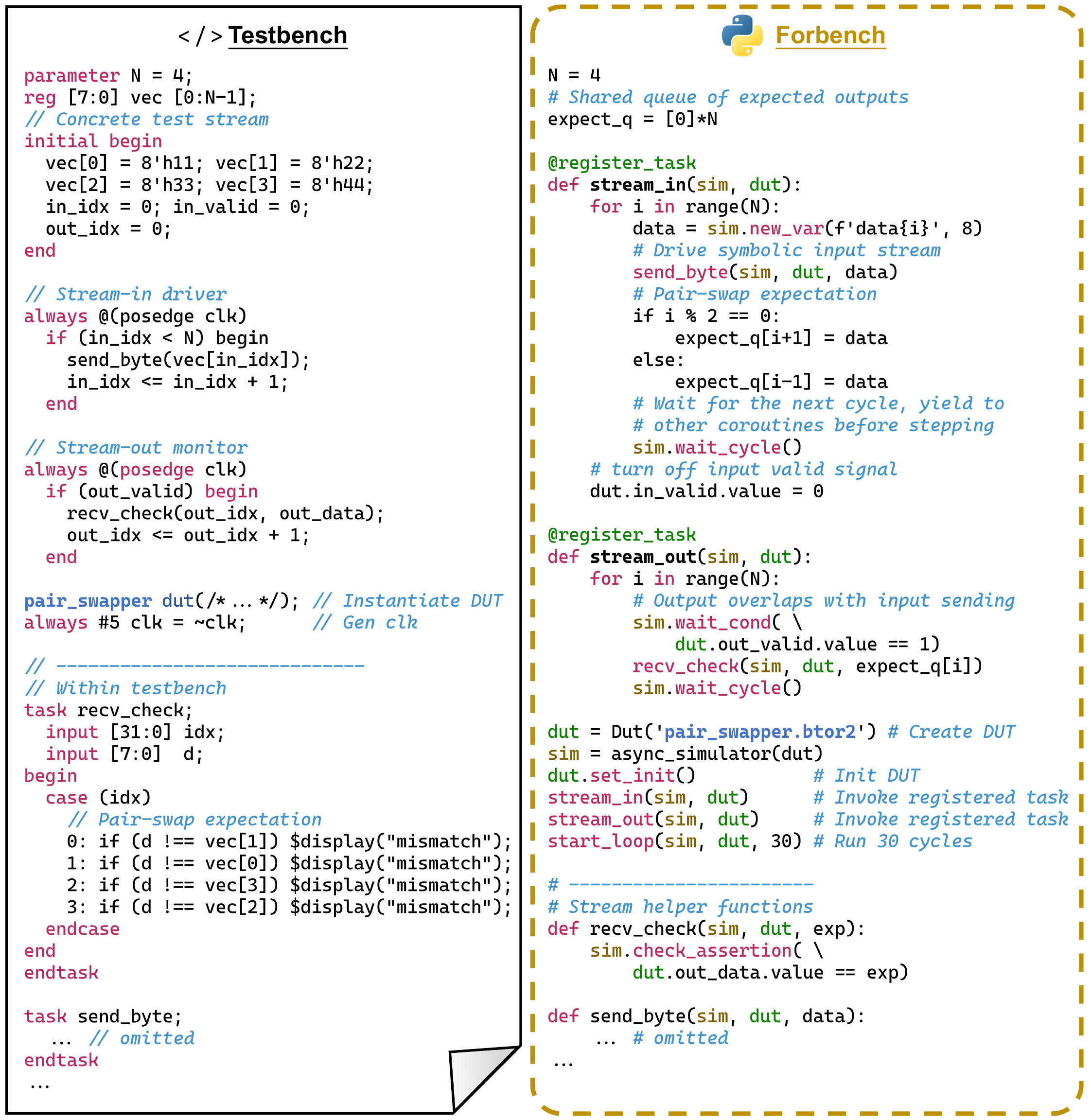}
  \caption{An example demonstrating verification of a streaming DUT with overlapped input/output sequences using either a traditional simulation testbench (left) or the symbolic testbench in \texttt{Forbench} (right). The two share a similar structure, with separate driver and monitor components and corresponding helper routines. In contrast to the significant structural gap between simulation and FPV assertions, Note how \texttt{Forbench} preserves the familiar testbench organization while enabling symbolic inputs (e.g., \texttt{data0} created using \texttt{sim.new\_var}) and assertion-based checking. }
  \label{fig:motivation}
  \vspace{-0.5cm}
\end{figure}

\smallskip
\noindent
\textbf{Initialization.} DUT can be initialized either by the reset condition coded in the Verilog (using  ``\textit{set\_init}'') or by a pure symbolic starting state (using ``\textit{free\_init}''). In the latter case,  all state variables in DUT are treated as free symbolic values. This allows the user to start the verification from an arbitrary intermediate state, which is desirable in modular inductive verification. In either case, the functions ``\textit{set\_init}'' and ``\textit{free\_init}'' also take an optional variable map as the argument to override the default variable assignment.


\smallskip
\noindent
\textbf{Simulation control.} 
The simulation advances in a cycle-accurate manner via the ``\textit{step}'' function, which provides fine-grained control over the application of stimuli at each cycle, where signal values can be accessed or modified using the form of \textit{dut.signal.value}.

\smallskip
\noindent
\textbf{Property checking.}
The users are free to take the value of signals in each cycle and construct logic formulas based these symbolic values.  The ``\textit{check\_assertion}'' function facilitates the verification of user-specified properties through SMT solving.
For assertions embedded in the RTL design,  ``\textit{check\_prop}'' can be used for a bounded verification of these properties.
There is another function ``\textit{set\_constraint}'' that allows users to impose assumptions on both input and state variables. These constraints can define fixed values, ranges, or complex symbolic relations, which restrict the state space and focus the analysis on expected behaviors.



\subsubsection{\textbf{Coroutine-based testbench}} 

Traditional hardware verification often relies on Verilog testbenches, where concurrency is achieved through parallel ``\textit{initial}'' and ``\textit{always}'' blocks. However, these testbenches are typically limited to simulating concrete, user-specified stimuli. While modern Python-based frameworks like \texttt{cocotb}~\cite{rosser2018cocotb} introduce coroutines to manage concurrent tasks, they still operate on a single, concrete execution trace at a time. 

In \texttt{Forbench}, we extend the idea of coroutine-based testbench to symbolic simulation. Our framework supports concurrent execution semantics similar to those in Verilog.
As shown by Fig.~\ref{fig:interface} ~\textcircled{b}, a Python function can be registered as a coroutine through the ``\textit{@register\_task}'' decorator. Within a testbench, there could be multiple coroutines. 
When the processing for a certain clock cycle has finished,  a coroutine could use 
``\textit{wait\_cond},'' 
``\textit{wait\_cycle},'' or ``\textit{wait\_task}'' to yield its execution to other coroutines that are still runnable within the same clock cycle. 

Specifically, the ``\textit{wait\_cond}'' pauses the execution of current coroutine to wait for a certain condition.
In the example of Fig.~\ref{fig:interface}~\textcircled{b}, the coroutine waits for the condition (``\textit{dut.valid.value == 1}'') to be met. \texttt{Forbench} formally checks the satisfiability of that condition under the current path constraints.
When the condition involves symbolic variables, both cases (the condition being met vs. unmet) could be possible, prompting the simulator to fork the execution into two coroutines, exploring both possibilities.
 In the given example, this creates a ``True'' path where the next assertion is checked immediately, and a ``False'' path where the clock steps forward and the condition will be re-evaluated in subsequent cycles.  Regarding other functions, ``\textit{wait\_cycle}'' pauses the current coroutine to wait for the simulation to advance by a specified number of cycles.  While one is waiting for the next cycle, others can proceed with their operations in the current cycle, ensuring proper interleaving of the testbench code among concurrent tasks. 
 Finally, the ``\textit{wait\_task}'' function is for synchronization, directing one coroutine to pause until another completes, managing dependencies among multiple tasks.\looseness=-1


The coordination among coroutines makes it very handy to work with overlapped I/O sequences on the DUT interface. Imagine that we are going to verify a DUT with a streaming I/O. As the whole input data stream is still sending in, some early output data may already start to stream out. If engineers use the active-stepping testbench with only a single point of execution, they would need to mix up the send-in and send-out processing code with the DUT stepping function, which results in messy code and is generally hard to maintain. Using coroutine-based testbench, the stream-in driver and the stream-out monitor could reside in separate coroutines, and with \texttt{Forbench}, the user can still enjoy the benefit of being symbolic, as shown by the code snippet in Fig.~\ref{fig:motivation}. This also addresses hurdle of dealing with those complex sequence-overlap problems in SVA composition when using FPV.\looseness=-1

\subsection{Symbolic States and Symbolic Simulation}
\label{3.2}

This subsection presents the formulation and the algorithms of symbolic simulation in \texttt{Forbench}.
We model RTL as a state transition system $\mathcal{TS} = (\mathcal{V}_s, \mathcal{V}_i, \mathcal{I}, \mathcal{T})$, where $\mathcal{V}_s $ and $ \mathcal{V}_i $ represent the sets of state and input variables, respectively. $\mathcal{I}$ specifies the initial conditions, and $\mathcal{T}$ defines the state transition relation.

\input{algorithm/algorithm}
At the core of our methodology is a symbolic simulation engine that explores the state space of $\mathcal{TS}$. The engine maintains a set of active symbolic states, $\mathbb{S}_{\text{active}}$, where each state $S^t$ at cycle $t$ is represented by a tuple $(\mathcal{M}_s^t, \Phi^t, \mathcal{X}^t)$. $\mathcal{M}_s$ maps each state variable $v \in \mathcal{V}_s$ to its symbolic expression and it gets updated per cycle to reflect system transitions. $\Phi$ is a set of symbolic formulas encoding constraints on the symbolic state, which include for example, user-specified assumptions and path constraints, where the latter accumulates path conditions associated with branching in the testbench.
$\mathcal{X}$ is a set of symbolic variables for unspecified input assignment or state variables with unspecified initial values. \texttt{Forbench} precisely preserves the relational information among these unspecified values. 
The handling of these values follows the approach introduced in the WASIM symbolic simulation framework~\cite{fang2023wasim}.
As an example, for an expression \textit{out = a+b}, where \texttt{a} is unspecified and \texttt{b} is assigned to be 1, it would be represented as \textit{X+1}, rather than a generic unknown value \textit{X} (here X is created as an element of $\mathcal{X}$).\looseness=-1


Symbolic simulation begins by constructing an initial symbolic state $S^0:(\mathcal{M}_s^0, \Phi^0, \mathcal{X}^0)$ based on $\mathcal{I}$ which is added to $\mathbb{S}_{\text{active}}$ (Lines 2-3 in Algorithm~\ref{alg:framework}).
The engine then iteratively processes each state $S^t$ from the active set (Line 5). 
Within each iteration, assertions are checked only when they are requested by the testbench in the current clock cycle (Lines 7-9), and potential forks are considered when a ``\textit{wait\_condition}'' is presented in the testbench (Line 10). Finally, it computes the next symbolic state $S^{t+1}$  by applying the transition relations $\mathcal{T}$ and adds it to the active set for exploration in the next cycle (Lines 11-18). 
For those input variables without specified assignments in the testbench, they will be mapped to fresh X-variables to capture all possible stimuli.
As long as there are no counterexample encountered,
this iterative process will continue, until there are no more clock stepping commands (for the active-stepping testbench) or when all coroutines terminate (for the coroutine-based testbench). A maximum clock cycle count $t_{max}$ can be provided to bound the symbolic simulation, ensuring termination.\looseness=-1

To account for the escalating complexity of symbolic expressions during symbolic simulation, we employ a set of techniques for  expression simplification. First, independence and constant analyses identify X-variables that are irrelevant to current transitions and substitute them with constant or simplified forms. Assumption-guided evaluation further resolves Boolean and bit-vector expressions whose outcomes can be determined under the current path conditions. 
In addition, unreachable cases in control structures like \textit{if-then-else} (ITE) are also removed.
%
%
Finally, to curb the growth of semantically equivalent expressions across states, we integrate the word-level sweeping technique~\cite{yang2025smtsweep}, which identifies and merges equivalent sub-expressions to reduce the complexity of symbolic expressions. 
Together, these simplification mechanisms keep symbolic expressions compact and tractable, ensuring that the symbolic simulation engine remains efficient as exploration depth increases.

\input{algorithm/coroutine_algorithm}


\subsection{Forking Coroutines}
\label{3.3}

As previously illustrated in Example~\textcircled{b} of Fig.~\ref{fig:interface}, coroutine forking is essential for exploring both branches of a symbolic condition. However, Python’s native coroutine system lacks the ability to replicate a coroutine’s execution state at runtime. Once a coroutine begins execution, its local variables, program counter, and internal stack evolve in-place and cannot be cloned, making it impossible to fork a running coroutine from an intermediate point.

To enable coroutine-based symbolic simulation, we develop a custom coroutine framework that supports fork-able coroutines. This framework leverages Python’s introspection and runtime evaluation capabilities to intercept, schedule, and manage the execution of coroutine bodies.
%
The \texttt{Forbench} framework then employs these fork-able coroutines to support concurrent explorations of different symbolic conditions.
When a coroutine encounters a conditional expression $E$ that could decide the execution path of the testbench, such as the one in a ``\textit{wait\_cond}'' statement, the algorithm determines whether to fork the current coroutine based on the outcome of SMT checks, as presented in Algorithm~\ref{alg:coroutine}.
 Given a symbolic state $S_1^t$ with its current path condition $\Phi^t$, the engine first checks the satisfiability of both the true path ($\Phi^t \land E$) and the false path ($\Phi^t \land \neg E$) using an SMT solver (Lines 2-3).  If only one of the two paths is satisfiable, the execution proceeds along that single feasible path, and the corresponding condition ($E$ or $\neg E$) is added to the path condition $\Phi^t$ (Lines 4-9). This step effectively prunes the impossible branch and prevents the exploration of an unreachable state.
Conversely, if both paths are satisfiable, it signifies a genuine symbolic branch point where both outcomes are possible (Line 10). In this case, the framework forks the execution by cloning the current state $S_1^t$ to create a new state $S_2^t$ (Line 11). The original state is updated to represent one branch by adding $\neg E$ to its path condition, while the new state represents the other by adding $E$ to its path condition (Lines 12-13). Each cloned state inherits the full context of its parent but proceeds with a distinct path condition. This forking mechanism allows the simulator to explore divergent execution paths concurrently, ensuring comprehensive coverage of all feasible design behaviors. 

Please note that in \texttt{Forbench}, \textbf{coroutine forking is driven by the conditions specified in the \textit{testbench}, rather than the branching conditions in the \textit{design} under verification}. This is a key distinction from prior symbolic execution approaches. In practice, the number of conditions in the testbench is typically much fewer than those in the DUT, which results in significantly fewer coroutines.
As will be shown by the experiments in Section~\ref{bug-finding-vs-symex}, fewer branchings lead to improved verification performance. Moreover, symbolic expressions are maximally reused across coroutines---\texttt{Forbench} retains only a single memory copy of shared sub-expressions in those cloned states, thereby minimizing the overhead associated with forking.

\vspace{-1.5em}
\subsection{State Merging to Cope with Path Explosion}
\label{3.4}
As discussed above, in general, symbolic simulation is less prone to the path explosion  problem compared to symbolic execution. That said, 
 in preparation of the extreme cases, \texttt{Forbench} also offers a mechanism for  symbolic state abstraction to merge the coroutines that are previously forked. Specifically, the simulator offers a function \texttt{sim.abstract} which takes a state predicate $\mathcal{P}$ as the argument. For every coroutine that arrives at the same point of this function call, the simulator will check if the state predicate $\mathcal{P}$ is an over-approximation of the current symbolic state $S^t : \left(\mathcal{M}_s^t, \Phi^t, \mathcal{X}^t \right)$.
 In case $\mathcal{P}$ is an over-approximation, we also say that the symbolic state $S^t$ satisfies/models $\mathcal{P}$ (namely, $S^t \models \mathcal{P}$).
 This checking is achieved by querying the validity of the following formula through SMT solving:
$\forall \mathcal{X}^t . \Phi^t\rightarrow\mathcal{P}|_{\mathcal{M}^t} $, where $\mathcal{P}|_{\mathcal{M}^t}$ means substituting the state variables in $\mathcal{P}$ according to the map $\mathcal{M}^t$.
Those  states satisfying $\mathcal{P}$ can be over-approximated by a single abstract state $S_a: \left(\mathcal{M}_a, \mathcal{P}|_{\mathcal{M}_a}, \mathcal{X}_a\right)$, where $\mathcal{M}_a$ maps each state variable to a newly-created X-variable. These  X-variables collectively satisfy the constraint $\mathcal{P}$, and are recorded in the set $\mathcal{X}_a$.

In this way, a set of symbolic states $\{S^t\}$ can all be replaced by a single abstract state $S_a$, and the coroutines associated with those symbolic states will be merged. As $\mathcal{P}$ is checked to be an over-approximation during merging, this operation will not lose the soundness of the analysis. Human experience and techniques like assertion/property mining~\cite{vasudevan2010goldmine} can be applied to construct such state predicates. 
Note that this functionality is just for the rare cases, and is not needed for the experiments presented in Section~\ref{sec:experiment}.


\vspace{-0.3cm}

\subsection{Application Notes}
\label{3.5}

Due to the space limitation, we are not able to present the full range of \texttt{Forbench} use cases in this paper, but still we would like to highlight several representative scenarios that can help better illustrate the flexibility and practical value of \texttt{Forbench}.

\vspace{0.2em}
\noindent
\textbf{Symbolic tandem simulation}.
A golden reference model is commonly required to validate the correctness of a DUT. In \texttt{Forbench}, this reference can be symbolically simulated in parallel with the DUT, effectively enabling symbolic differential testing. Leveraging prior work such as SeaHorn~\cite{gurfinkel2015seahorn}, even software models written in C/C++ can be compiled into formal representations (e.g., constrained Horn clauses~\cite{gurfinkel2022program}), thereby enabling cross-language symbolic co-simulation between hardware and software models.

\vspace{0.2em}
\noindent
\textbf{Modular inductive verification}.
Symbolic simulation in \texttt{Forbench} naturally supports symbolic initial states, allowing the verifier to start from arbitrary (possibly nondeterministic) states rather than being restricted to the reset state. This provides significantly greater flexibility and coverage: DUT behaviors that manifest only after deep execution (e.g., millions of cycles after reset) can be analyzed without requiring prohibitively long simulations. Similar strategies have proven effective in prior work such as \cite{fadiheh2018symbolic}.

\vspace{0.2em}
\noindent
\textbf{Backward symbolic simulation}.
Although this paper focuses primarily on forward symbolic execution, \texttt{Forbench} also supports \emph{backward} symbolic simulation. Given a symbolic state satisfying certain constraints, the engine can compute the symbolic representation of predecessor states. This capability parallels the backward reasoning techniques equipped in GSTE~\cite{1218209} and enables reasoning about behaviors that span varying numbers of cycles.

%% file: algorithm/algorithm.tex
\begin{algorithm}[t]
\caption{: Symbolic Simulation in \texttt{Forbench}}
\label{alg:framework}

\resizebox{0.95\columnwidth}{!}{%
\begin{minipage}{\columnwidth}

\begin{algorithmic}[1]
\Require Transition system $\mathcal{TS} = (\mathcal{V}_s, \mathcal{V}_i, \mathcal{I}, \mathcal{T})$, max cycle $t_{max}$
\Ensure Property verification result: $\{\textsc{Safe}, \textsc{Unsafe}\}$
\State \textbf{\textit{Step1: construct the initial symbolic state}}
\State $\mathcal{M}_s^0 \gets \{v \mapsto \text{init}_v \mid \forall v \in \mathcal{V}_s\}$ \Comment{initial state variable map}

\State $\mathbb{S}_{\text{active}} \gets \{(\mathcal{M}_s^0, \Phi^0, \mathcal{X}^0)\}$ \Comment{initial symbolic state set}

\State \textit{\textbf{Step2: clock stepping}} 
\For {\textbf{each} $S^t = (\mathcal{M}_s^t, \Phi^t, \mathcal{X}^t) \in \mathbb{S}_{\text{active}}$ } 

    \State \textbf{if} $t>t_{max}$ \, \textbf{then} \, break
    \If {$\Call{CheckAssertion}{S^t}$ is \textsc{Sat}} 
        \State \Return \textsc{Unsafe} \Comment{Early stop when assertion violated}
    \EndIf
    \State $\mathbb{S}_{\text{active}} \gets \mathbb{S}_{\text{active}} \cup \{\Call{CheckBranch}{S^{t}} \}$  \Comment{in Algorithm~\ref{alg:coroutine}}
    \State $\mathcal{M}_{\text{spec}}^t \gets \Call{GetSpecInput}{t}$ \Comment{specified input-value map}
    \State $U_i^t \gets \mathcal{V}_i \setminus \mathrm{domain}(\mathcal{M}_{\text{spec}}^t)$  \Comment{unspecified inputs}
    \State ${\mathcal{X}}_{\text{new}} \gets \Call{Create}{U_i^t}$ \Comment{create X-variables}
    \State $\mathcal{X}^{t+1} \gets \mathcal{X}^{t} \cup {\mathcal{X}}_{\text{new}}$ 

    \State $\mathcal{M}_i^t \gets \mathcal{M}_{\text{spec}}^t \cup \{v \mapsto {\mathcal{X}}_v \mid v \in U_i^t, {\mathcal{X}}_v \in {\mathcal{X}}_{\text{new}}\}$ 
    \State $\mathcal{M}_s^{t+1} \gets \{v \mapsto \mathcal{T}(\mathcal{M}_s^t, \mathcal{M}_i^t) \mid v \in \mathcal{V}_s\}$
    \State $S^{t+1} \gets (\mathcal{M}_s^{t+1}, \Phi^{t+1}, \mathcal{X}^{t+1})$ \Comment{execute one step}
    \State $ \mathbb{S}_{active} \gets \mathbb{S}_{active} \cup \{S^{t+1}\} $ 
\EndFor 
\State \Return \textsc{Safe} \Comment{safe uptill the explored cycles}


\end{algorithmic}
\end{minipage}
}

\end{algorithm}

%% file: algorithm/coroutine_algorithm.tex
\begin{algorithm}[t]
\caption{: \textsc{CheckBranch} Subprocedure}
\label{alg:coroutine}

\resizebox{0.95\columnwidth}{!}{%
\begin{minipage}{\columnwidth}
\begin{algorithmic}[1]

\Require Symbolic state $S_1^t = (\mathcal{M}_s^t, \Phi^t, \mathcal{X}^t)$
\Ensure At most two symbolic states, denoted as $S_1^t$ and $S_2^t$.

\If{current task waits on condition $E$} 
    \State $R_{true}$ =  \Call{CheckSat}{${\Phi^{t} \land E}$}
    \State $R_{false}$ = \Call{CheckSat}{${\Phi^{t} \land \neg E}$}
    \If {$R_{false}$ is \textsc{Sat}, $R_{true}$ is \textsc{Unsat}}
        \State $S_1^{t}.\Phi^{t} \gets \Phi^{t} \cup \{\neg E\} $ \Comment{accumulate path condition}
        \State \Return $S_1^t$
    \ElsIf{$R_{true}$ is \textsc{Sat}, $R_{false}$ is \textsc{Unsat}}
        \State $S_1^{t}.\Phi^{t} \gets \Phi^{t} \cup \{E\} $ \Comment{accumulate path condition}
        \State \Return $S_1^t$
    \ElsIf{both $R_{true}$, $R_{false}$ are \textsc{Sat}}
        \State $S_2^t \gets \Call{Clone}{S_1^t}$ \Comment{clone the symbolic state}
        \State ${S_1^{t}.\Phi^{t}} \gets \Phi^{t} \cup \{\neg E\} $ \Comment{accumulate condition for $S_1^t$}
        \State $S_2^{t}.{\Phi}^{t} \gets \Phi^{t} \cup \{E\} $ \Comment{accumulate condition for $S_2^t$}
        \State \Return $S_1^t, S_2^t$
    \EndIf \Comment{$R_{true}$ and $R_{false}$ cannot be both \textsc{Unsat}}
\EndIf
\State \Return $S_1^t$ 
\end{algorithmic}
\end{minipage}
}
\end{algorithm}

    

%% file: sections/experiment.tex
\label{sec:experiment}

This section evaluates \texttt{Forbench} in terms of (i) simulation performance, and efficiency in (ii) DUT branch covering and (iii) bug-finding.\looseness=-1


\subsection{Experiment Setup}
The \texttt{Forbench} simulation engine is implemented in C++, utilizing SMT-Switch~\cite{mann2021smt} to interface with the \texttt{Bitwuzla} SMT solver~\cite{niemetz2023bitwuzla}. Hardware designs in Verilog are first compiled into the BTOR2 model~\cite{niemetz2018btor2} via Yosys~\cite{wolf2013yosys} and are then loaded into the simulator.
\texttt{Forbench} exposes a Python API through Boost.Python that interacts with the symbolic simulator. 
Experiments are run on a server with Ubuntu 20.04.4 LTS, dual Intel Xeon Platinum 8375C processors and 256GB RAM.

\subsection{Experiment 1: Simulation Performance}

\noindent
\textbf{Baselines.}
We compare \texttt{Forbench} with representative engines from three different verification styles: two concrete RTL simulators, Icarus Verilog (Iverilog)~\cite{williams2002icarus} and Verilator~\cite{snyder2004verilator}; \texttt{Wasim}~\cite{fang2023wasim}, a word-level symbolic simulator; and \texttt{VossII}~\cite{Seger2020VossII}, a symbolic trajectory evaluation (STE)-based tool. 

\vspace{0.3em}
\noindent
\textbf{Benchmark.} For this experiment, we evaluate on a set of five RTL designs with different scales to better reflect runtime scalability. These include complex algorithmic and processor-scale RTL designs, including the AES implementation and several open-source processor cores.
These designs contain on the order of $10^4$--$10^5$ gates, and therefore stress the scalability of symbolic simulation on realistic hardware blocks rather than on toy examples.


\vspace{0.3em}
\noindent
\textbf{Results.}
The results are reported in Table~\ref{tab:sim_time}. 
As expected, \texttt{iverilog} and \texttt{verilator} achieve the lowest wall-clock time, since each run executes only a single concrete trace for one concrete input assignment. Their runtime therefore reflects the cost of simulating one valuation, rather than reasoning about the full input space. 
\texttt{VossII}, as an STE tool, is in principle more powerful than concrete simulation because it symbolically reasons over sets of bounded trajectories. However, this strength comes with substantial BDD-based reasoning overhead, which makes STE much less scalable on datapath-heavy and data-intensive designs. In our experiments, \texttt{VossII} could not complete even a single cycle for AES or Rocket within the 1-hour time limit. By contrast, both \texttt{Wasim} and \texttt{Forbench} use SMT-based symbolic simulation, so a single run can reason about symbolic inputs and achieve exhaustive coverage under the stated bound. Among these symbolic approaches, \texttt{Forbench} is the most practical one across all five designs: it consistently outperforms \texttt{Wasim}, owning to the SMT simplification techniques mentioned near the end of Section~\ref{3.2} and also the efficiency of C++ implementation (whereas the core of symbolic simulation in \texttt{Wasim} is written in Python). \texttt{Forbench} also runs faster than \texttt{VossII} on all five designs, and still provides full bounded symbolic coverage rather than the result of only one concrete execution.\looseness=-1

\input{table/sim_time}

\input{table/coverage}

\subsection{Experiment 2: Branch Covering Efficiency}

\noindent
\textbf{Baselines.} 
We further compare \texttt{Forbench} with several state-of-the-art formal~\cite{mann2021pono,ebmc2022} and coverage-guided concolic testing~\cite{zheng2023stsearch,zheng2025hotfv} tools using branch coverage as a metric.


\vspace{0.3em}
\noindent
\textbf{Benchmark.} 
To ensure apples-to-apples comparisons with these prior works, especially the more recent ones: STSearch~\cite{zheng2023stsearch} and Hot-FV~\cite{zheng2025hotfv}, we take the benchmark designs from these prior works, which can be organized into small and large ones.
The small benchmarks originally come from the ITC'99 suite~\cite{davidson1999characteristics}, a standardized collection of digital circuits widely used to evaluate test-generation techniques; the instances used in our evaluation include \texttt{b01}, \texttt{b06}, \texttt{b10}, \texttt{b11}, \texttt{b12}, and \texttt{b14}. The large benchmarks are processor-scale RTL designs from OR1200 and Rocket. OR1200 is an open-source 32-bit RISC processor for embedded applications, featuring a configurable architecture, a five-stage pipeline, separate instruction and data caches, and DSP support~\cite{openrisc_or1200_2020}. In Table~\ref{tab:branch_cov}, \texttt{Exception}, \texttt{ICache}, and \texttt{DCache} denote the OR1200 modules for exception handling, instruction-cache control, and data-cache control, respectively, while \texttt{or1200} denotes the complete processor. Rocket is a five-stage in-order scalar core implementing the RV64GC RISC-V instruction set and generated from the Rocket Chip framework~\cite{asanovic2016rocket}. The design is originally written in Chisel and can produce RTLs of different sizes through configuration. We therefore include two RocketTile configurations, \texttt{rocketTile\_tiny} and \texttt{rocketTile\_small}, to evaluate how the proposed approach behaves as processor complexity increases.

\vspace{0.3em}
\noindent
\textbf{Results.}
Table~\ref{tab:branch_cov} reports the runtime and coverage of all methods. \texttt{Forbench} achieves branch coverage comparable to the best results from existing tools, while offering significantly better runtime performance. Compared to random testing, \texttt{Forbench} is far more targeted: it reaches higher coverage on benchmarks such as \texttt{b14}, \texttt{Exception}, and \texttt{ICache}, and triggers deep or corner-case branches in substantially less time (e.g., \texttt{b11} and \texttt{DCache}). Relative to other formal and concolic approaches, \texttt{Forbench} matches their coverage and is generally more efficient on the nontrivial cases. For the larger-scale \texttt{or1200} and \texttt{rocketTile} benchmarks, \texttt{Forbench} further shortens the runtime while maintaining the same coverage level, which is closely tied to our strategy for controlling branch explosion during symbolic exploration.\looseness=-1

\subsection{Experiment 3: Bug-Finding Efficiency}
\label{bug-finding-vs-symex}

\noindent
\textbf{Baseline.}
Bug finding provides a complementary perspective beyond branch coverage metric alone. Bruns \emph{et al.}~\cite{bruns2023processor} showed that symbolic execution can effectively uncover both injected and real bugs in a RISC-V processor, demonstrating strong bug-hunting capability. In this experiment, we compare \texttt{Forbench} with this symbolic execution approach in both the bug-finding capability and efficiency.\looseness=-1

\input{table/bug_finding}

\vspace{0.3em}
\noindent
\textbf{Benchmark.} 
The experiment setup follows the public available code repository associated with the work~\cite{bruns2023processor}. The design under verification is the micro-riscv processor~\cite{Ahmadi-Pour_MicroRV32} 
supporting the RV32I instruction set. It is originally designed in SpinalHDL, and then compiled into Verilog for verification.
As the public code repository does not contain the RTL code with injected errors, to reproduce the bug-finding results of the prior work, we follow the descriptions in its text to re-create the same ten injected bugs (E0--E9), covering representative classes of processor design errors. Specifically, E0 introduces an incorrect \texttt{SLLI} decode; E1--E2 model don't-care errors in \texttt{SRLI} decoding; E3 and E4 inject stuck-at faults into the \texttt{ADDI} and \texttt{SUB} datapaths, respectively; E5 corrupts the PC update logic for \texttt{JAL}; E6 makes \texttt{BNE} behave as \texttt{BEQ}; and E7--E9 target load semantics, including an endianness error in \texttt{LBU}, missing sign extension in \texttt{LB}, and a faulty \texttt{LW} implementation that loads only 16 bits.
For symbolic execution, we set the instruction limit to 1, which is the most efficient configuration as presented in the paper~\cite{bruns2023processor}.
We then also applied \texttt{Forbench} to designs with the same injected bugs to compare for the bug-finding capability/efficiency.


\vspace{0.3em}
\noindent
\textbf{Results.}
Table~\ref{tab:bug_finding} reports detailed quantitative results for all 10 injected bugs. Both approaches successfully detect all injected bugs. Compared with the existing symbolic execution method, \texttt{Forbench} substantially reduces the time for bug detection in this micro-riscv processor.  This is because it only needs to explore much fewer branches in the testbench compared to the branches in the DUT, supporting  our claim---although \texttt{Forbench} forks coroutines upon dual-outcome symbolic branching conditions encountered in the testbench, it is still more efficient compared to traditional symbolic execution, which branches according to conditions in DUT.


\looseness=-1

%% file: table/sim_time.tex
\setcounter{table}{1}
\begin{table}[t]
  \centering
  \small
  \setlength{\tabcolsep}{3pt}
  \resizebox{\linewidth}{!}{
    \begin{threeparttable}
      \caption{Simulation time (wall-clock) comparison (in seconds)}
      \vspace{-0.2cm}
      \label{tab:sim_time}
      \begin{tabular}{l r r | c c | c c c}
        \toprule
          \makecell[c]{Designs} &
          \makecell[c]{\#Gates} &
          \makecell[c]{Verilog LOC} &
          \makecell[c]{\textbf{Iverilog}} &
          \makecell[c]{\textbf{Verilator}} &
          \makecell[c]{\textbf{Wasim}} &
          \makecell[c]{\textbf{VossII}$^{\dagger}$} &
          \makecell[c]{\textbf{Forbench}} \\
        \midrule
        aes~\cite{pasalaNarayan_AES_Verilog}   & 195561 & 425    & 0.86 & 0.06 & 5.98    & >3600  & \cellcolor{lightblue}1.60 \\
        rocket~\cite{asanovic2016rocket}      & 26963  & 11,572 & 0.02 & 0.01 & 187.42  & >3600  & \cellcolor{lightblue}10.81 \\
        piccolo~\cite{piccolo2018}            & 51878  & 6,078  & 0.11 & 0.02 & 726.66  & N/A  & \cellcolor{lightblue}63.54 \\
        flute~\cite{flute2018}                & 72980  & 10,212 & 0.04 & 0.01 & 1048.30 & N/A & \cellcolor{lightblue}125.16 \\
        ridecore~\cite{ridecore2016}          & 377324 & 6,678  & 0.28 & 0.37 & 1127.65 & N/A & \cellcolor{lightblue}79.83 \\
        \bottomrule
      \end{tabular}
      \begin{tablenotes}[flushleft]
        \item $\dagger$: N/A for several designs due to Yosys compatibility issues in VossII 4.0; time out (>3600s) on the remaining designs.
        \item Note: \texttt{iverilog} and \texttt{verilator} report measured wall-clock time for exact concrete simulation. \texttt{VossII} reports measured wall-clock time for symbolic trajectory evaluation, while \texttt{Wasim} and \texttt{Forbench} report measured wall-clock time for symbolic simulation; all five use a simulation bound of 100 cycles.
      \end{tablenotes}
    \end{threeparttable}
  }
  \vspace{-2em}
\end{table}

%% file: table/coverage.tex
\setcounter{table}{2}
\begin{table*}[tbp]
  \centering
  \resizebox{0.95\textwidth}{!}{%
    \begin{minipage}{\textwidth}
      \small
      \setlength{\tabcolsep}{2pt}
      \begin{threeparttable}
       \vspace{-1.2em}
        \caption{Branch coverage and runtime comparison}
    \vspace{-1em}
        \label{tab:branch_cov}
        \begin{tabular}{l cc cccccc cccccc}
      \toprule
      \multirow{2}{*}{\textbf{Benchmarks\runtimemark}} &
        \multicolumn{2}{c}{\textbf{Characteristics}} &
        \multicolumn{6}{c}{\textbf{Branch Coverage (\%)}} &
        \multicolumn{6}{c}{\textbf{Time (s)}} \\
      \cmidrule(lr){2-3}\cmidrule(lr){4-9}\cmidrule(lr){10-15}
      & \textbf{Branches} & \textbf{Lines} &
        \textbf{Random} & \textbf{Pono} & \textbf{EBMC} & \textbf{STSearch} & \textbf{Hot-FV}  & \textbf{Forbench} &
        \textbf{Random}\tnote{\ddag} & \textbf{Pono} & \textbf{EBMC} & \textbf{STSearch} & \textbf{Hot-FV}   & \textbf{Forbench}\\
      \midrule
      b01       & 26  & 159  & 100.00 & 100.00 & 100.00 & 100.00  & 100.00 & \textbf{\cellcolor{lightblue}100.00 } & 0.02 & 1.66 & 0.74 & \textbf{0.01} & 1.47 & \cellcolor{lightblue}0.23 \\
      b06       & 24  & 166  & 95.83  & 95.83  & 95.83  & 95.83   & 95.83  & \textbf{\cellcolor{lightblue}95.83} & \textbf{0.01} & 1.65 & 0.61 & \textbf{0.01} & 1.48 & \cellcolor{lightblue}0.13 \\
      b10       & 41  & 252  & 100.00 & 100.00 & 100.00 & 100.00  & 100.00 & \textbf{\cellcolor{lightblue}100.00} & 0.11 & 2.43 & 4.92 & \textbf{0.01} & 1.50 & \cellcolor{lightblue}0.55 \\
      b11       & 32  & 203  & 96.88  & 96.88  & 96.88  & 96.88   & 96.88  & \textbf{\cellcolor{lightblue}96.88 }  & 9.57 & 2.77 & 4.96 & 4.54 & 2.38 & \textbf{\cellcolor{lightblue}0.57} \\
      b14       & 193 & 1141 & 97.44  & 98.96  & 98.96  & 97.93   & 98.96  & \textbf{\cellcolor{lightblue}98.96}  & \textbf{1.08} & 7.21 & 145.95 & 97.93 & 4.74 & \cellcolor{lightblue}1.22 \\
      Exception & 47  & 468  & 95.74  & 100.00 & 100.00 & 97.87   & 100.00 & \textbf{\cellcolor{lightblue}100.00} & 0.85 & 2.30 & 19.77 & 7.17 & 1.58 & \textbf{\cellcolor{lightblue}0.24} \\
      DCache    & 46  & 377  & 97.83  & 97.83  & 97.83  & 91.30   & 97.83  & \textbf{\cellcolor{lightblue}97.83}  & 47.76 & 2.30 & 2.42 & 4.38 & 1.53 & \textbf{\cellcolor{lightblue}0.46} \\
      ICache    & 26  & 180  & 84.62  & 96.15  & 96.15  & 96.15   & 96.15  & \textbf{\cellcolor{lightblue}96.15}  & 0.16 & 1.65 & 2.42 & 6.90 & 1.50 & \textbf{\cellcolor{lightblue}0.07} \\
      b12              & 112 & 737 & 32.15 & 88.39 & 86.61 & 34.82 & 89.29 & \textbf{\cellcolor{lightblue} 89.29 }  & 91.03 & 5684.03 & 4428.85 & 3068.73 & 3890.04 & \textbf{\cellcolor{lightblue}1816.30} \\
      or1200           & 782 & 14262 & 58.82 & 94.63 & N/A$^{\dagger}$ & 59.85 & 94.63 & \textbf{\cellcolor{lightblue}94.63} & 11058.42 & 144.86 & N/A & 8983.78 & 222.99 & \textbf{\cellcolor{lightblue}83.64} \\
      rocketTile\_tiny & 1522 & 57028 & 47.83 & 85.81 & N/A$^{\dagger}$ & 61.83 & 85.81 & \textbf{\cellcolor{lightblue}85.81} & 16361.94 & 7818.87 & N/A & 9100.73 & 5468.8 & \textbf{\cellcolor{lightblue}4106.56} \\
      rocketTile\_small & 2198 & 75496 & 39.44 & 79.16 & N/A$^{\dagger}$ &  48.91 & 79.16 & \textbf{\cellcolor{lightblue}79.16} & 10004.28 & 29408.5 & N/A & 11451.03 & 10831.04 & \textbf{\cellcolor{lightblue}9437.96} \\
      \bottomrule
        \end{tabular}

        \begin{tablenotes}[flushleft]
          \footnotesize
          \item [\(\lozenge\)]: Benchmarks are selected from open-source processor designs~\cite{openrisc_or1200_2020,asanovic2016rocket} and the ITC'99 benchmark suite~\cite{davidson1999characteristics}.
          \item [\ddag]: For random-based approaches, we report the time at which the branch coverage first reaches its maximum within the 30,000-second limit. 
          \item [\dag]: EBMC produced exceptions and did not finish this benchmark.
        \end{tablenotes}
      \end{threeparttable}
    \end{minipage}%
  }
  \vspace{-1.2em}
\end{table*}

%% file: table/bug_finding.tex
\begin{table}[tp]
  \centering
  \small
  \setlength{\tabcolsep}{2.5pt}
  \renewcommand{\arraystretch}{1.12}
  \caption{Bug-finding capability and efficiency 
  }
  \vspace{-0.3cm}
  \label{tab:bug_finding}
  \resizebox{0.85\linewidth}{!}{%
    \begin{tabular}{l c r r r r c r c}
      \toprule
      \multirow{2}{*}{\makecell[c]{\textbf{Error}}}
      & \multicolumn{5}{c}{\textbf{Symbolic Execution}}
      & \multicolumn{3}{c}{\textbf{Forbench}} \\
      \cmidrule(lr){2-6}\cmidrule(lr){7-9}
      & \makecell[c]{\textbf{Result}}
      & \makecell[c]{\textbf{\#Exec.}\\\textbf{Instr.}}
      & \makecell[c]{\textbf{Time}\\\textbf{(s)}}
      & \makecell[c]{\textbf{Partial}\\\textbf{Paths}}
      & \makecell[c]{\textbf{Paths}}
      & \makecell[c]{\textbf{Result}}
      & \makecell[c]{\textbf{Time}\\\textbf{(s)}}
      & \makecell[c]{\textbf{Branches}} \\
      \midrule
      E0 & \checkmark & 26,141,972 & 1,526.39 & 1,396 & 141 & \cellcolor{lightblue}\checkmark & \cellcolor{lightblue}0.85 & \cellcolor{lightblue}2 \\
      E1 & \checkmark & 13,947,436 & 1,563.74 & 1,477 & 59  & \cellcolor{lightblue}\checkmark & \cellcolor{lightblue}0.87 & \cellcolor{lightblue}2 \\
      E2 & \checkmark & 12,553,774 & 1,459.77 & 1,478 & 53  & \cellcolor{lightblue}\checkmark & \cellcolor{lightblue}0.69 & \cellcolor{lightblue}2 \\
      E3 & \checkmark & 6,648,904  & 767.02   & 1,351 & 21  & \cellcolor{lightblue}\checkmark & \cellcolor{lightblue}1.10 & \cellcolor{lightblue}2 \\
      E4 & \checkmark & 6,571,348  & 794.26   & 1,351 & 21  & \cellcolor{lightblue}\checkmark & \cellcolor{lightblue}0.87 & \cellcolor{lightblue}2 \\
      E5 & \checkmark & 9,590,444  & 2,769.18 & 1,476 & 42  & \cellcolor{lightblue}\checkmark & \cellcolor{lightblue}0.84 & \cellcolor{lightblue}2 \\
      E6 & \checkmark & 6,187,314  & 784.28   & 1,349 & 21  & \cellcolor{lightblue}\checkmark & \cellcolor{lightblue}0.73 & \cellcolor{lightblue}2 \\
      E7 & \checkmark & 8,660,026  & 915.09   & 1,484 & 36  & \cellcolor{lightblue}\checkmark & \cellcolor{lightblue}3.68 & \cellcolor{lightblue}3 \\
      E8 & \checkmark & 8,732,876  & 864.43   & 1,484 & 36  & \cellcolor{lightblue}\checkmark & \cellcolor{lightblue}4.08 & \cellcolor{lightblue}3 \\
      E9 & \checkmark & 2,492,982  & 201.08   & 226   & 0   & \cellcolor{lightblue}\checkmark & \cellcolor{lightblue}3.66 & \cellcolor{lightblue}3 \\
      \midrule
      Sum    & 10/10 & 101,527,076 & 11,645.24 & 13,072 & 430 & 10/10 & 17.37 & 23 \\
      Median & --    & 8,696,451   & 889.76     & 1,436  & 36  & --    & 0.87  & 2 \\
      \bottomrule
    \end{tabular}%
  }
\vspace{-2.5em}
\end{table}

%% file: sections/conclusion.tex
\label{sec:conclusion}
This paper presents \texttt{Forbench}, a novel formal testbench paradigm that preserves the familiar testbench styles while adding symbolic reasoning capabilities for easier use and better verification performance. The supporting framework is available at \url{https://github.com/hkustgz-zhang-lab/Forbench}.\looseness=-1

